\documentclass[conference]{IEEEtran}
\IEEEoverridecommandlockouts

\usepackage{cite}
\usepackage{amsmath,amssymb,amsfonts}
\usepackage{algorithmic}
\usepackage{graphicx}
\usepackage{textcomp}
\usepackage{xcolor}
\usepackage{float} 
\usepackage{listings}
\usepackage[T1]{fontenc}
\usepackage[utf8]{inputenc}
\usepackage{url}
\usepackage{booktabs}
\usepackage{fancyhdr}
\usepackage{hyperref}
\usepackage{setspace}

\lstdefinelanguage{json}{
    basicstyle=\ttfamily\footnotesize,
    numbers=left,
    numberstyle=\tiny\color{gray},
    stepnumber=1,
    numbersep=5pt,
    showstringspaces=false,
    breaklines=true,
    frame=tb,
    literate=
     *{0}{{{\color{blue}0}}}{1}
      {1}{{{\color{blue}1}}}{1}
      {2}{{{\color{blue}2}}}{1}
      {3}{{{\color{blue}3}}}{1}
      {4}{{{\color{blue}4}}}{1}
      {5}{{{\color{blue}5}}}{1}
      {6}{{{\color{blue}6}}}{1}
      {7}{{{\color{blue}7}}}{1}
      {8}{{{\color{blue}8}}}{1}
      {9}{{{\color{blue}9}}}{1}
      {:}{{{\color{red}:}}}{1}
      {,}{{{\color{red},}}}{1}
      {\{}{{{\color{red}\{}}}{1}
      {\}}{{{\color{red}\}}}}{1}
      {[}{{{\color{red}[}}}{1}
      {]}{{{\color{red}]}}}{1},
}
\begin{document}

\title{\mbox{Blockchain-Based Decentralized Domain Name System}}

\author{
\IEEEauthorblockN{Guang Yang, Peter Trinh, Alma Nkemla, \\
Amuru Serikyaku, Edward Tatchim, Osman Sharaf}
\IEEEauthorblockA{\{guangyang19, trinhp, almankemla, amuru, edwardtatchim, sharafosman\}@berkeley.edu\\
University of California, Berkeley}
}

\maketitle

\begin{abstract}
The current Domain Name System (DNS) infrastructure faces critical vulnerabilities including poisoning attacks, censorship mechanisms, and centralized points of failure that compromise internet freedom and security. Recent incidents such as the APT Group StormBamboo DNS poisoning attacks on ISP customers demonstrate the urgent need for resilient alternatives. This paper presents a novel blockchain-based Decentralized Domain Name System (DDNS). We designed a specialized Proof-of-Work blockchain to maximize support for DNS-related protocols and achieve node decentralization. The system integrates our blockchain with IPFS for distributed storage, implements cryptographic primitives for end-to-end trust signatures, achieving Never Trust, Always Verify zero-trust verification. Our implementation achieves 15-second domain record propagation times, supports 20 standard DNS record types, and provides perpetual free .ddns domains. The system has been deployed across distributed infrastructure in San Jose, Los Angeles, and Orange County, demonstrating practical scalability and resistance to traditional DNS manipulation techniques. Performance evaluation shows the system can handle up to Max Theor. TPS 1,111.1 tx/s (minimal transactions) / Max Theor. TPS 266.7 tx/s (regular transactions) for domain operations while maintaining sub-second query resolution through intelligent caching mechanisms.
\end{abstract}


\section{Introduction}

\subsection{Problem Statement}

The modern internet's Domain Name System (DNS) represents a critical infrastructure vulnerability that undermines both security and freedom of information. Two primary categories of threats have emerged as systemic challenges:

\textbf{DNS Security Vulnerabilities:} The centralized architecture of traditional DNS systems creates attractive targets for sophisticated attacks. Recent evidence includes the APT Group StormBamboo attacks, which compromised ISP-level DNS infrastructure to redirect legitimate traffic to malicious endpoints~\cite{stormbamboo2024}. These poisoning attacks exploit the inherent trust relationships in hierarchical DNS resolution, demonstrating how centralized control points become systemic weaknesses~\cite{berger2019wrinkle,wei2021whac}.

\textbf{Censorship and Access Restrictions:} Authoritarian regimes and restrictive governments increasingly employ DNS-based censorship as a mechanism for information control. Large-scale DNS record manipulation and selective blocking of domain resolution violate fundamental principles of information freedom and democratic access to knowledge. This systematic interference with DNS infrastructure represents a technological assault on human rights to free expression and access to information.

The mathematical formulation of these problems can be expressed as single points of failure in the DNS resolution chain:

\begin{equation}
P_{failure} = 1 - \prod_{i=1}^{n} (1 - p_i)
\end{equation}

where $p_i$ represents the failure probability of the $i$-th centralized component in the DNS hierarchy, and $n$ is the number of critical control points.

\subsection{Motivation}

Traditional DNS systems operate under a trust model that is susceptible to single point of failure problems and prone to security and availability risks. The hierarchical structure creates dependencies on centralized authorities (root servers, top-level domain registrars, ISPs) that can be compromised, coerced, or corrupted. This centralization enables attacks that violate the CIA (Confidentiality, Integrity, Availability) security principles:

\begin{itemize}
    \item \textbf{Single Point of Failure Attacks:} Compromising availability (Availability), where compromise of authoritative servers can affect millions of domains simultaneously
    \item \textbf{State-Level Censorship:} Compromising confidentiality (Confidentiality), where governments can mandate DNS filtering at ISP or national levels
    \item \textbf{Commercial Manipulation:} Compromising availability (Availability), where domain registrars can unilaterally suspend or transfer domains
    \item \textbf{Data Integrity Violations:} Compromising integrity (Integrity), where DNS responses lack cryptographic verification, enabling man-in-the-middle attacks
\end{itemize}

The proliferation of these attacks necessitates a paradigm shift toward cryptographically secured, decentralized domain name zero-trust resolution that ensures confidentiality and integrity. This paradigm eliminates central points of control, thereby ensuring performance and availability.

\subsection{Our Contribution}

This paper presents a comprehensive blockchain-based decentralized DNS system (hereinafter referred to as DDNS) that addresses these fundamental limitations through:

\begin{enumerate}
    \item \textbf{DDNS Blockchain Infrastructure:} The project codename is Phicoin, representing an acronym for Proof of Work High-performance Infrastructure, aimed at building a fully decentralized blockchain network using PoW mechanism. A purpose-built Proof-of-Work blockchain optimized for domain asset management using modified UTXO model with enhanced transaction throughput (4MB blocks, 15-second intervals)
    
    \item \textbf{Cryptographic Trust Chain:} End-to-end verification using elliptic curve digital signatures (ECDSA) for domain registration and modification, eliminating reliance on third-party trust
    
    \item \textbf{Distributed Storage Integration:} IPFS-based domain control file storage with content-addressable hashing for tamper-evident record management
    
    \item \textbf{Anti-Censorship Architecture:} Mesh network communication patterns and distributed resolver infrastructure that circumvents traditional blocking mechanisms
    
    \item \textbf{Cross-chain Integration:} We developed cross-chain bridges capable of connecting the DDNS blockchain to other smart contract-enabled blockchains such as Solana and Ethereum. We have deployed smart contracts on Solana, enabling the 15 million monthly active users (MAU) of Solana to directly utilize DDNS services for domain registration and updates without downloading node clients.
    
    \item \textbf{Edge DDNS Resolution Services:} We developed edge DDNS servers that enable DDNS resolution within organizations without relying on public DDNS services, while maintaining compatibility with traditional DNS resolution. This requires only deploying edge DDNS services within the organization and configuring them as the organization's network DNS servers.
    
    \item \textbf{Decentralized Web Protocol (D-WEB):} We implemented a decentralized D-WEB protocol based on DDNS TXT records, which leverages IPFS's capability to resolve static directories, using TXT records to resolve website IPFS hashes, enabling decentralized archiving/access of traditional websites.
\end{enumerate}

\section{Related Work}

Existing decentralized naming systems have made significant contributions to addressing DNS centralization, but each suffers from fundamental limitations that prevent widespread adoption~\cite{yang2024ddns}. Recent surveys on blockchain consensus mechanisms provide comprehensive frameworks for evaluating different approaches to decentralized systems~\cite{xiao2020survey,bano2017consensus}.

\textbf{Ethereum Name Service (ENS):} Utilizes Ethereum's smart contract infrastructure for .eth domain management~\cite{ens2024}. While innovative, ENS faces scalability constraints due to Ethereum's throughput limitations (Max Theor. TPS 119.1 tx/s) and high transaction costs (gas fees often exceeding \$50 per operation). Additionally, ENS domains are not compatible with traditional DNS infrastructure, limiting their utility~\cite{wood2014ethereum}.

\textbf{Namecoin:} The first blockchain-based naming system, forked from Bitcoin to support .bit domains~\cite{namecoin2014}. Namecoin suffers from slow block times (10 minutes), limited throughput, and lack of modern DNS record type support. The system's security relies on merge-mining with Bitcoin, creating potential centralization risks~\cite{antonopoulos2014mastering}.

\textbf{Handshake:} Implements a novel approach using proof-of-work to manage top-level domain auctions~\cite{handshake2021}. However, Handshake focuses primarily on TLD ownership rather than practical DNS resolution, and its auction mechanism creates significant barriers to entry for users.

While these systems represent important advances in decentralized naming, they each exhibit fundamental trade-offs between security, scalability, and usability that limit their practical deployment. Recent research in blockchain consensus mechanisms has shown that achieving optimal balance between these properties requires careful design of the underlying consensus protocol.

Our system advances the state-of-the-art by combining:
- High throughput blockchain infrastructure (Max Theor. TPS 1,111.1 tx/s vs. Max Theor. TPS 119.1 tx/s for Ethereum)
- Universal DNS compatibility (supports all standard record types)
- Zero-cost operation for .ddns domains
- Production-ready resolver infrastructure

\begin{table}[H]
  \centering
  \caption{Comparison of Decentralized DNS Solutions}
  \resizebox{0.48\textwidth}{!}{%
  \begin{tabular}{lccccc}
  \toprule
  \textbf{Feature} & \textbf{DDNS} & \textbf{ENS} & \textbf{Namecoin} & \textbf{Handshake} & \textbf{Traditional DNS} \\
  \midrule
  Decentralized     & Yes            & Partial        & Yes            & Yes            & No              \\
  DNS Compatible    & Yes            & No             & Limited        & Limited        & Yes             \\
  Block Time        & 15s            & 12s            & 10min          & 10min          & N/A             \\
  Transaction Cost  & Free           & \$10-50        & \$0.01         & \$1-10         & \$10-100/year   \\
  Throughput (TPS)  & 1,111.1/266.7  & 119.1          & 7              & 7              & 250~\cite{cloudflare_performance}       \\
  Record Types      & 20             & Limited        & Limited        & Limited        & 20~\cite{cloudflare_dns} \\
  Censorship Resist & High           & Medium         & High           & High           & Low             \\
  \bottomrule
  \end{tabular}%
  }
\end{table}

% \textbf{Note:} The DDNS project aims to become foundational infrastructure in the Web3 decentralized domain. Following network stabilization, we will release V3 version with further optimized block times, providing faster transaction speeds to support increased DNS update requests.

\section{System Architecture}

\subsection{High-Level Design}

The DDNS system implements a layered architecture that separates concerns while maintaining cryptographic security guarantees throughout the stack. The design follows the principle of \textit{cryptographic minimalism}, where trust assumptions are explicitly modeled and minimized.

\begin{figure}[H]
\centering
\includegraphics[width=0.48\textwidth]{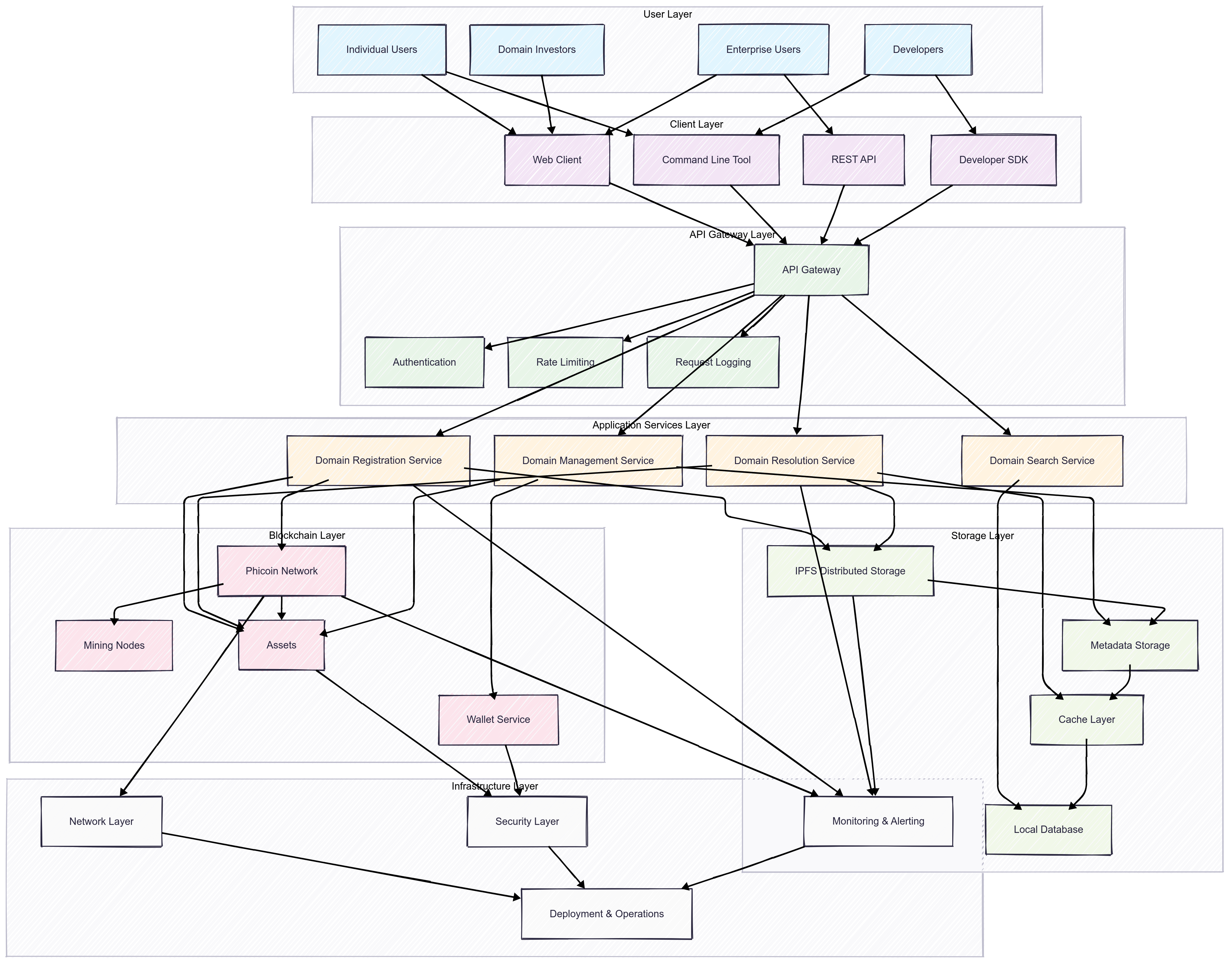}
\caption{DDNS System High-Level Architecture}
\label{fig:architecture}
\end{figure}

The architecture consists of six primary layers:

\textbf{User Layer:} Supports diverse stakeholders including individual users, enterprises, developers, and domain investors with varying technical requirements and economic models.

\textbf{Client Layer:} Provides multiple interfaces (Web UI, CLI tools, REST APIs, SDKs) for different integration patterns and user preferences.

\textbf{API Gateway Layer:} Implements authentication, rate limiting, and request logging with horizontal scaling capabilities.

\textbf{Application Services Layer:} Core business logic for domain registration, resolution, management, and search with high-availability design.

\textbf{Blockchain Layer:} DDNS network providing cryptographic consensus, asset management, wallet services, and mining infrastructure.

\textbf{Storage Layer:} Distributed storage using IPFS, metadata management, intelligent caching, and local database optimization.

\subsection{Functional Components}

The system architecture comprises multiple specialized functional components that work together to provide comprehensive domain management and resolution services.

\begin{figure}[H]
\centering
\includegraphics[width=0.48\textwidth]{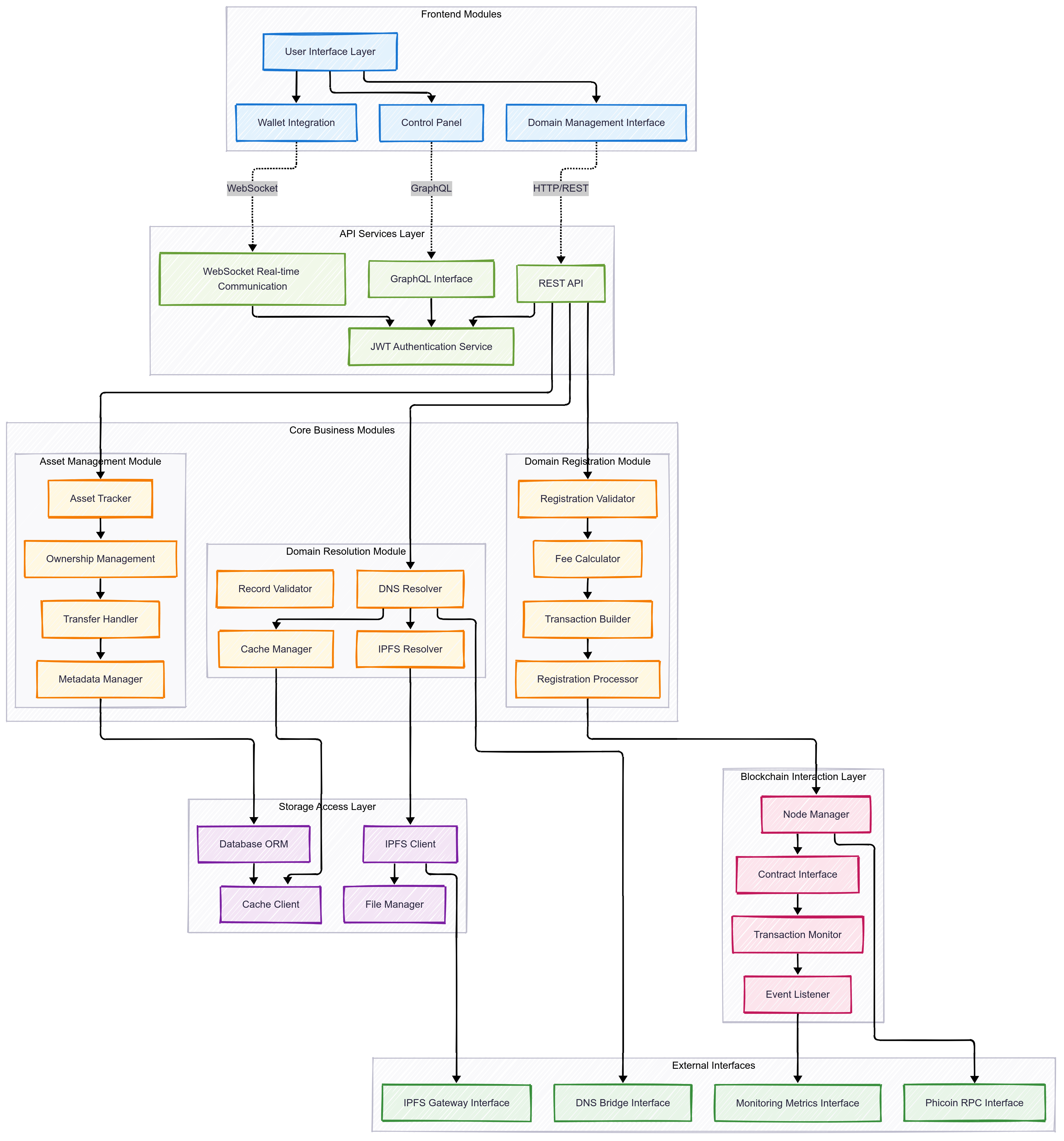}
\caption{DDNS Functional Components and Module Interactions}
\label{fig:functional_components}
\end{figure}

The functional architecture includes core business modules (Asset Management, Domain Registration), domain resolution modules (DNS Resolver, Cache Manager, IPFS Resolver), blockchain integration components (Node Manager, Contract Interface), and storage access layers (Database ORM, IPFS Client, File Manager). The frontend modules provide user interfaces through REST APIs, WebSocket real-time communication, and GraphQL interfaces for advanced querying capabilities.

\subsection{Use Case Analysis}

The system supports diverse user personas with varying technical expertise and usage patterns, from individual users seeking simple domain registration to enterprise users requiring bulk domain management capabilities.

\begin{figure}[H]
\centering
\includegraphics[width=0.48\textwidth]{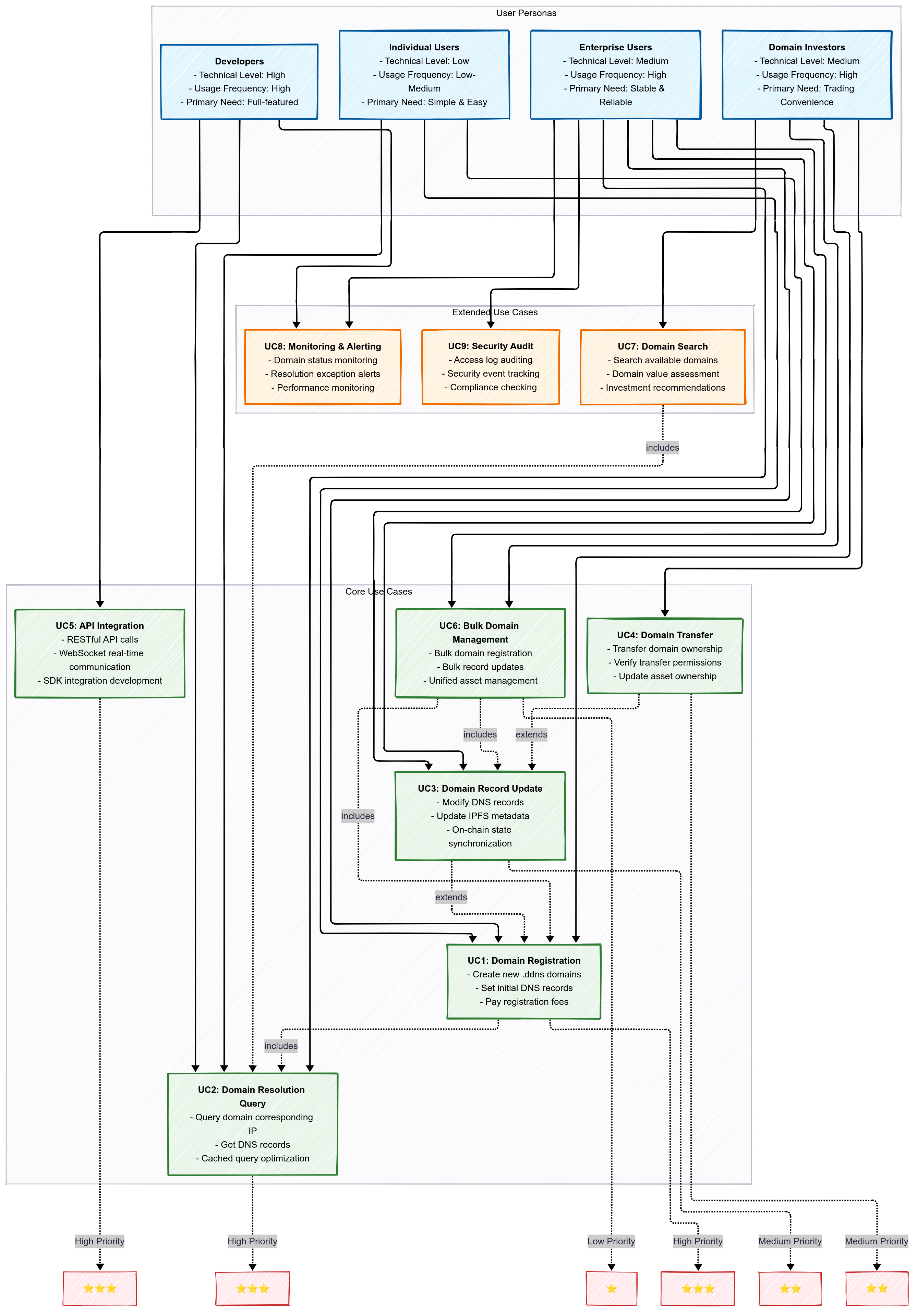}
\caption{DDNS Use Case Diagram and User Interactions}
\label{fig:use_cases}
\end{figure}

The use case diagram illustrates core functionalities including domain registration, record updates, domain resolution queries, bulk domain management, domain transfers, and monitoring/alerting services. Each use case is prioritized based on user needs and technical complexity, with high-priority cases including domain registration and resolution queries that form the foundation of the system.

\subsection{Technology Stack}

The DDNS implementation leverages a modern, scalable technology stack designed for high-performance blockchain and distributed systems operations.

\begin{figure}[H]
\centering
\includegraphics[width=0.48\textwidth]{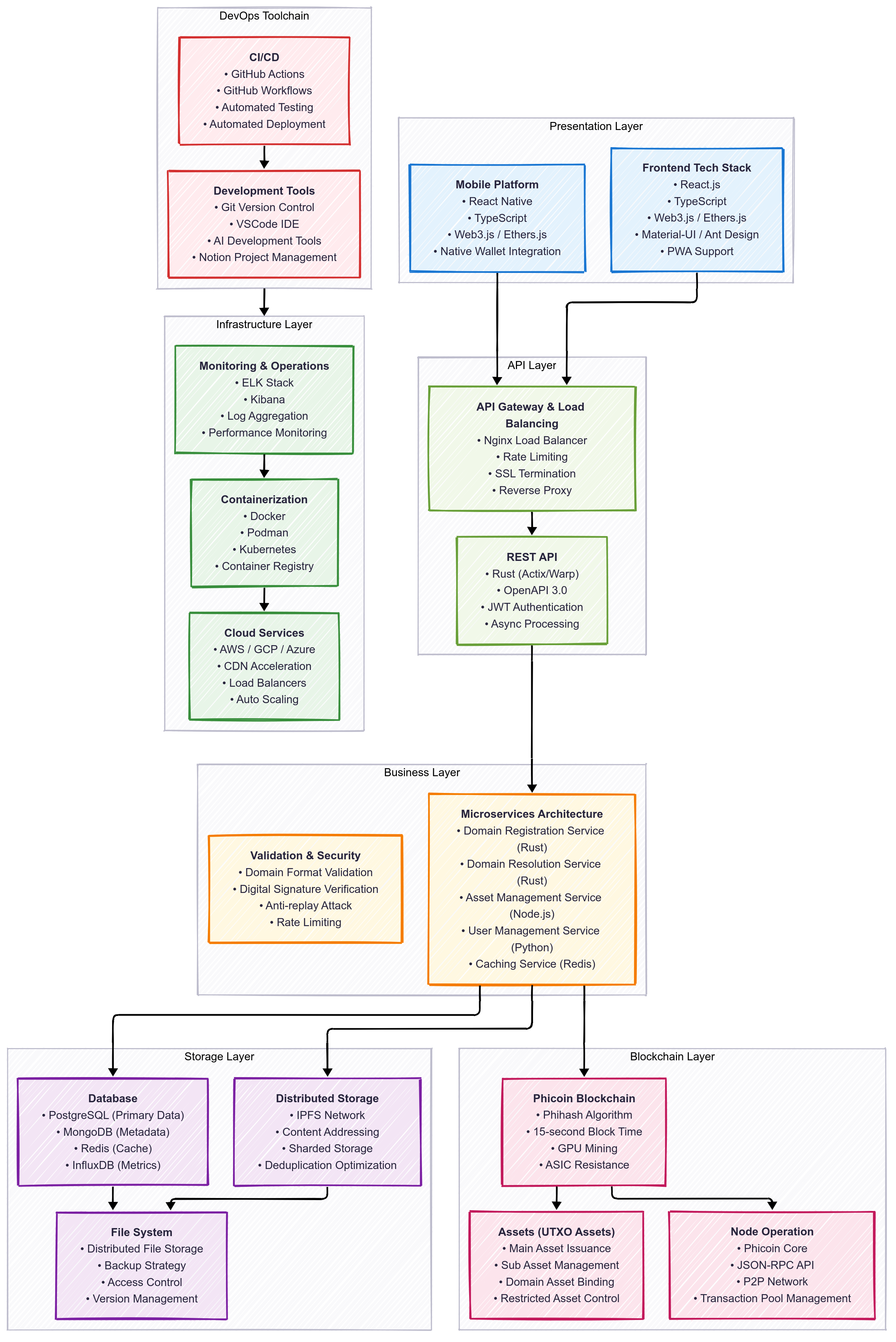}
\caption{DDNS Technology Stack and Infrastructure Components}
\label{fig:tech_stack}
\end{figure}

The technology stack spans multiple layers: the presentation layer utilizes React.js for frontend development and React Native for mobile platforms; the API layer implements load balancing with Nginx and RESTful services; the business layer employs microservices architecture with domain-specific services in Rust and Node.js; the storage layer combines distributed storage (IPFS), traditional databases (PostgreSQL, MongoDB), and Redis for caching; the blockchain layer features the custom DDNS blockchain with UTXO asset management; and the infrastructure layer provides containerization with Docker/Kubernetes, monitoring with ELK stack, and DevOps automation.

\subsection{Cryptographic Primitives}

The system employs well-established cryptographic primitives to ensure security:

\textbf{Digital Signatures:} Domain operations utilize ECDSA with secp256k1 curve (identical to Bitcoin) for signing transactions. The signature verification process follows:

\begin{equation}
\text{Verify}(m, \sigma, pk) = e(\sigma, G) \stackrel{?}{=} e(H(m) + r \cdot pk, G)
\end{equation}

where $m$ is the message (domain operation), $\sigma$ is the signature, $pk$ is the public key, and $H$ is SHA-256 hash function.

\textbf{Content Addressing:} IPFS uses SHA-256 hash function for content addressing:

\begin{equation}
\begin{split}
\text{IPFS\_Hash} &= \text{Base58}(\text{SHA256}( \\
&\quad \text{ProtoBuf}(\text{Domain\_Record})))
\end{split}
\end{equation}

This ensures tamper-evident storage where any modification to domain records results in a different hash value.

\subsection{DDNS Blockchain Infrastructure}

The DDNS blockchain represents a purpose-built blockchain optimized for domain name service requirements. Key technical specifications include:

\textbf{Consensus Algorithm:} Proof-of-Work using PhihashV2, an ASIC-resistant algorithm designed to prevent network hash power centralization that leads to centralization risks~\cite{garay2015bitcoin,nakamoto2008bitcoin}. Additionally, we optimized the DAG structure to enable cache files to run on integrated graphics cards, allowing tens of millions of devices worldwide equipped with integrated graphics to join the mining network at any time, enabling broader network participation and larger-scale decentralized networks. This approach addresses the well-documented mining centralization concerns in PoW systems~\cite{eyal2014majority,gencer2018decentralization}.

\textbf{Block Parameters:}
- Block time: 15 seconds (optimized for DNS update responsiveness)
- Block size: 4MB Weight Units (WU) = 4,000,000 WU (enabling high transaction throughput)
- Difficulty adjustment: Enhanced Dark Gravity Wave algorithm for rapid response

\textbf{Transaction Throughput:} The system achieves theoretical maximum throughput using Weight Units calculation method~\cite{nakamoto2008bitcoin}:

General TPS calculation formula (applicable only to DDNS Blockchain: 4M WU):
\begin{equation}
TPS = \frac{4,000,000}{\text{Average Weight Units per Transaction}} \div 15
\end{equation}

Different transaction types yield the following results:
\begin{enumerate}
\item Minimal transactions (bare transfers, typically 60 bytes = 240 WU):
\begin{equation}
TPS_{minimal} = \frac{4,000,000}{240} \div 15 \approx 1,111.1
\end{equation}

\item Regular transactions ($\approx$ 250 bytes = 1,000 WU):
\begin{equation}
TPS_{regular} = \frac{4,000,000}{1,000} \div 15 = 266.7
\end{equation}
\end{enumerate}

\textbf{Final Results Summary (4M WU):}
\begin{table}[H]
\centering
\caption{Transaction Types and Theoretical Maximum TPS}
\label{tab:tps_summary}
\footnotesize
\setlength{\tabcolsep}{4pt}
\begin{tabular}{@{}lcc@{}}
\toprule
\textbf{Transaction Type} & \textbf{Average Size (WU)} & \textbf{TPS (Theoretical Max)} \\
\midrule
Minimal Transaction & 240 WU & $\approx$ 1,111.1 \\
Regular Transaction & 1,000 WU & $\approx$ 266.7 \\
\bottomrule
\end{tabular}
\end{table}

\vspace{0.5cm}

\textbf{Network Security:} Empirical analysis of mainnet performance from block heights 92,594 to 143,669 shows orphan rate of 0.0176\% (9 orphaned blocks out of 51,075 total)~\cite{phicoin_orphans}, demonstrating network stability.

\subsection{Domain Asset Model}

The system implements a modified UTXO model specifically designed for domain asset management. Each domain is represented as a unique asset with the following properties:

\begin{lstlisting}[language=json,caption={Domain Asset Structure}]
{
  "asset_name": "DDNS/EXAMPLE",
  "quantity": 1,
  "units": 1,
  "reissuable": false,
  "has_ipfs": true,
  "ipfs_hash": "QmX7M8RxZ...",
  "owner_address": "PhiCoinAddress123..."
}
\end{lstlisting}

\textbf{Asset Naming Convention:} Domains follow hierarchical naming: \texttt{ROOT\_TLD/DOMAIN\_NAME} where ROOT\_TLD represents the top-level domain asset (e.g., "DDNS") and DOMAIN\_NAME represents the specific domain.

\textbf{Economic Model:} Domain registration requires minimal fees (0.1 PHI $\approx$ \$0.00001) with no recurring costs, implementing true digital asset ownership rather than lease-based models used by traditional DNS. Additionally, we subsidize the gas fees for domain registration. For .DDNS domain registration, we cover all gas fees, thereby achieving permanently free domain services.

\subsection{IPFS Integration and Domain Control Files}

Domain records are stored in JSON-formatted control files on IPFS, enabling flexible schema evolution and comprehensive DNS record type support~\cite{ipfs_whitepaper}. The control file structure follows RFC-compliant specifications:

\begin{lstlisting}[language=json,caption={Domain Control File Example}]
{
  "version": "2.0",
  "domain": "example.ddns",
  "records": {
    "@": {
      "A": [{"address": "192.168.1.100", "ttl": 3600}]
    },
    "www": {
      "CNAME": [{"target": "example.ddns", "ttl": 3600}]
    },
    "mail": {
      "MX": [{"server": "mail.example.ddns", "priority": 10}]
    }
  }
}
\end{lstlisting}

This design provides several advantages:
- \textbf{Schema Flexibility:} JSON format allows arbitrary record types without blockchain protocol changes
- \textbf{Efficient Storage:} Only IPFS hash stored on-chain, enabling large record sets without blockchain bloat
- \textbf{Content Verification:} IPFS content addressing ensures data integrity

\subsection{Supported DNS Record Types}

The DDNS system provides comprehensive support for 20 different DNS record types, ensuring compatibility with modern internet infrastructure requirements and enabling diverse use cases from simple web hosting to complex service architectures. We have implemented 76 types of domain resolution and control files enumerated in RFC documents. However, we also referenced Cloudflare's protocol resolution settings and selected the 20 most commonly used domain resolution records. The aforementioned control file code and scripts are open-sourced in the ddnsd service, allowing users to configure according to their specific needs.

\textbf{Core Address Records:}
\begin{itemize}
\item \textbf{A Records~\cite{rfc1035}:} IPv4 address mapping for standard web services and applications
\item \textbf{AAAA Records~\cite{rfc3596}:} IPv6 address mapping supporting next-generation internet protocols
\item \textbf{CNAME Records~\cite{rfc1035}:} Canonical name aliases enabling flexible domain management and CDN integration
\end{itemize}

\textbf{Mail and Communication Records:}
\begin{itemize}
\item \textbf{MX Records~\cite{rfc1035}:} Mail exchange server specifications with priority-based routing
\item \textbf{TXT Records~\cite{rfc1035}:} Arbitrary text data for SPF, DKIM, domain verification, and custom metadata
\item \textbf{SPF Records~\cite{rfc7208}:} Sender Policy Framework for email authentication and anti-spam protection
\item \textbf{DKIM Records~\cite{rfc6376}:} DomainKeys Identified Mail cryptographic signatures for email integrity
\item \textbf{DMARC Records~\cite{rfc7489}:} Domain-based Message Authentication for comprehensive email security policies
\end{itemize}

\textbf{Service Discovery Records:}
\begin{itemize}
\item \textbf{SRV Records~\cite{rfc2782}:} Service location records defining port and priority for specific services
\item \textbf{NS Records~\cite{rfc1035}:} Name server delegation for subdomain management and distributed authority
\item \textbf{PTR Records~\cite{rfc1035}:} Reverse DNS lookups enabling IP-to-domain resolution
\item \textbf{SOA Records~\cite{rfc1035}:} Start of Authority defining zone management parameters and refresh intervals
\end{itemize}

\textbf{Advanced and Specialized Records:}
\begin{itemize}
\item \textbf{CAA Records~\cite{rfc6844}:} Certificate Authority Authorization controlling SSL/TLS certificate issuance
\item \textbf{TLSA Records~\cite{rfc6698}:} Transport Layer Security Authentication for DNS-based certificate pinning
\item \textbf{SSHFP Records~\cite{rfc4255}:} SSH Key Fingerprints for secure shell authentication verification
\item \textbf{URI Records~\cite{rfc7553}:} Uniform Resource Identifier mapping for advanced service location
\item \textbf{NAPTR Records~\cite{rfc3403}:} Naming Authority Pointer for complex protocol transformations
\item \textbf{LOC Records~\cite{rfc1876}:} Geographic location information for physical server positioning
\item \textbf{HINFO Records~\cite{rfc1035}:} Host information describing system architecture and operating system
\item \textbf{RP Records~\cite{rfc1183}:} Responsible Person contact information for domain administration
\end{itemize}

\textbf{Record Type Validation:} Each record type implements RFC-compliant validation ensuring data integrity and standards compliance. The system performs real-time validation during record updates, preventing malformed entries and maintaining DNS protocol compatibility.

\textbf{Performance Optimization:} Record resolution is optimized through intelligent caching with type-specific TTL policies, reducing resolution latency for frequently accessed record types while maintaining accuracy for dynamic records.

\section{Implementation}

\subsection{Blockchain Infrastructure Components}

The DDNS network consists of multiple specialized components working in concert:

\textbf{Core Node Software:} Full blockchain nodes implementing the DDNS protocol, maintaining complete transaction history, and participating in consensus.

\textbf{Mining Infrastructure:} Distributed mining pools supporting the PhihashV2 algorithm, with specialized mining software optimized for GPU hardware~\cite{phihashminer_github}.

\textbf{Network Discovery:} Seeder servers providing initial peer discovery and network health monitoring.

\textbf{DDNSD Public DDNS Resolution Servers:} Distributed public resolution infrastructure providing high-availability domain name resolution services.

\textbf{DDoH Public DDNS over HTTPS Resolution Servers:} Secure DNS resolution services implementing DNS-over-HTTPS protocol for enhanced privacy and censorship resistance.

\textbf{Cross-Chain Bridge:} Smart contract infrastructure on Solana enabling PHI token trading and liquidity provision, creating economic incentives for network participation.

\textbf{Sustainable Infrastructure:} The DDNS network operates on environmentally sustainable infrastructure, including solar-powered data centers that provide carbon-neutral blockchain operations. Our San Jose facility demonstrates the feasibility of renewable energy-powered PoW cryptocurrency block perpetual generation node operations.

\begin{figure}[H]
\centering
\includegraphics[width=0.5\textwidth]{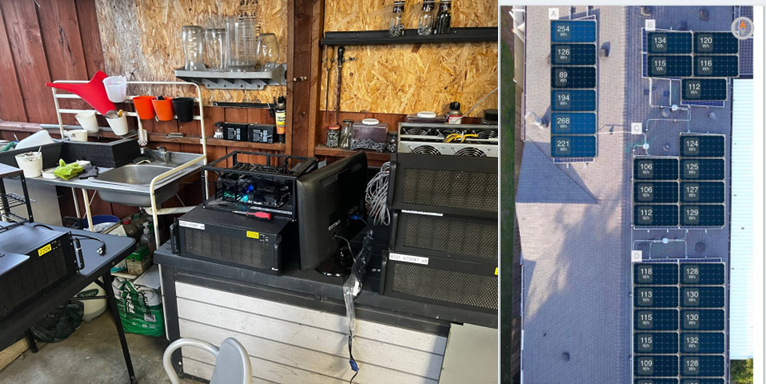}
\caption{Solar-Powered Data Center Infrastructure in San Jose}
\label{fig:solar_infrastructure}
\end{figure}

The solar infrastructure includes high-efficiency photovoltaic panels, battery storage systems, and optimized cooling solutions that enable 24/7 blockchain node operation with minimal environmental impact. This sustainable approach to blockchain infrastructure addresses growing concerns about cryptocurrency energy consumption while maintaining network security and performance.

\subsection{Domain Registration and Modification Process}

The domain lifecycle follows a cryptographically secured process:

\textbf{Registration Flow:}
\begin{enumerate}
    \item User generates key pair $(sk, pk)$ where $sk$ is private key and $pk = sk \cdot G$ is corresponding public key
    \item Create domain control file with initial DNS records
    \item Upload control file to IPFS, obtaining hash $h_{ipfs}$
    \item Construct blockchain transaction $tx = \{domain\_name, h_{ipfs}, pk\}$
    \item Sign transaction: $\sigma = \text{ECDSA\_Sign}(sk, H(tx))$
    \item Broadcast signed transaction to DDNS network
    \item Miners validate signature and include in next block
\end{enumerate}

\textbf{Modification Flow:} Domain updates follow identical process but reference existing asset, ensuring only authorized private key holder can modify records.

\begin{figure}[H]
\centering
\includegraphics[width=0.5\textwidth]{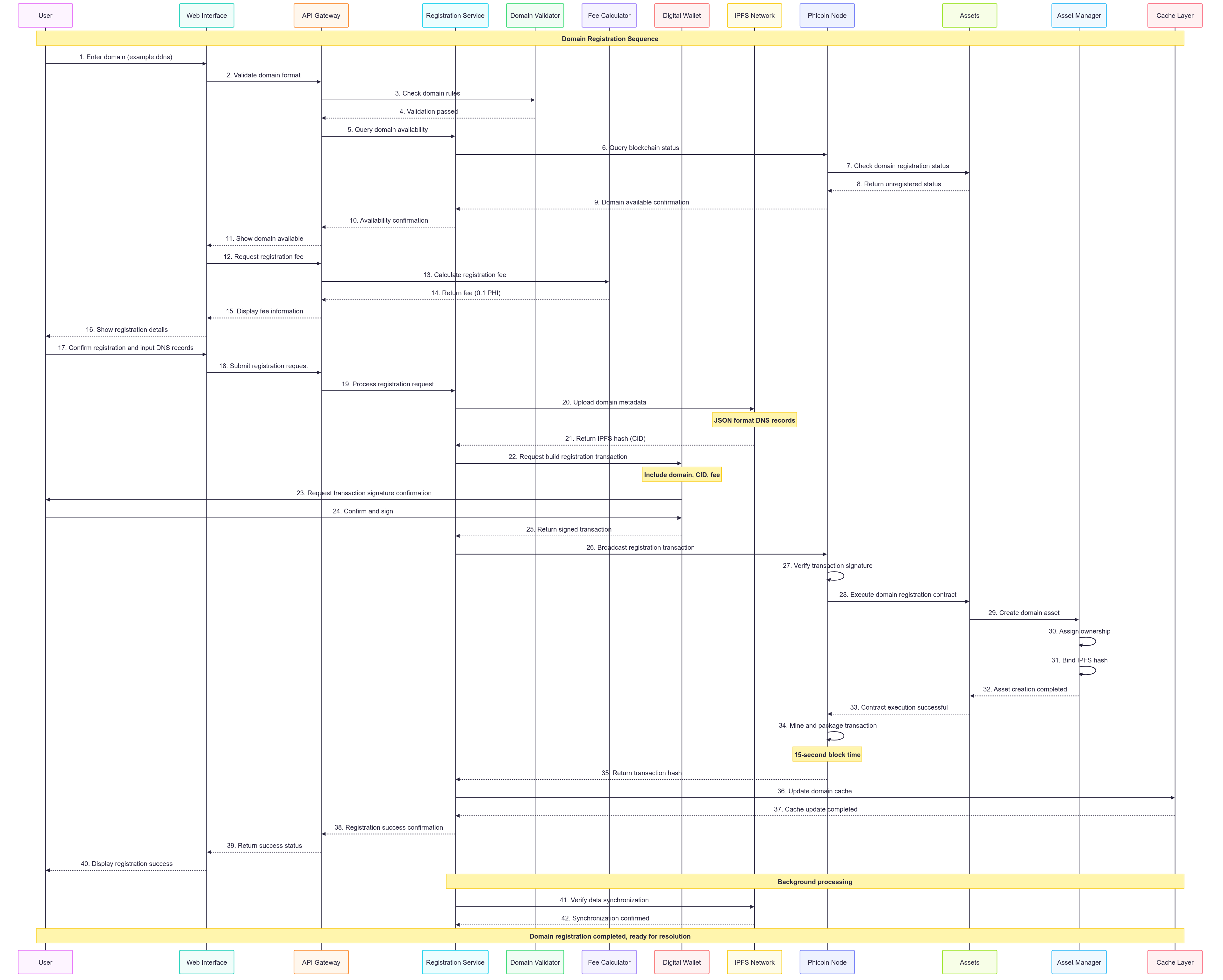}
\caption{Domain Registration Sequence Diagram}
\label{fig:domain_registration}
\end{figure}

The domain registration sequence demonstrates the complete end-to-end flow from user initiation through blockchain confirmation. The process involves multiple system components including the Web Interface, API Gateway, Registration Service, Domain Validator, Fee Calculator, Digital Wallet, IPFS Network, DDNS Node, and Asset Manager. Key steps include domain validation, fee calculation, IPFS metadata upload, blockchain transaction creation and signing, and final asset creation with ownership assignment. The sequence emphasizes the cryptographic security at each step and the 15-second block time for rapid confirmation.

\subsection{DNS Resolution Process}

The resolution system implements a hybrid approach combining blockchain verification with traditional DNS performance requirements:

\begin{figure}[H]
\centering
\includegraphics[width=0.5\textwidth]{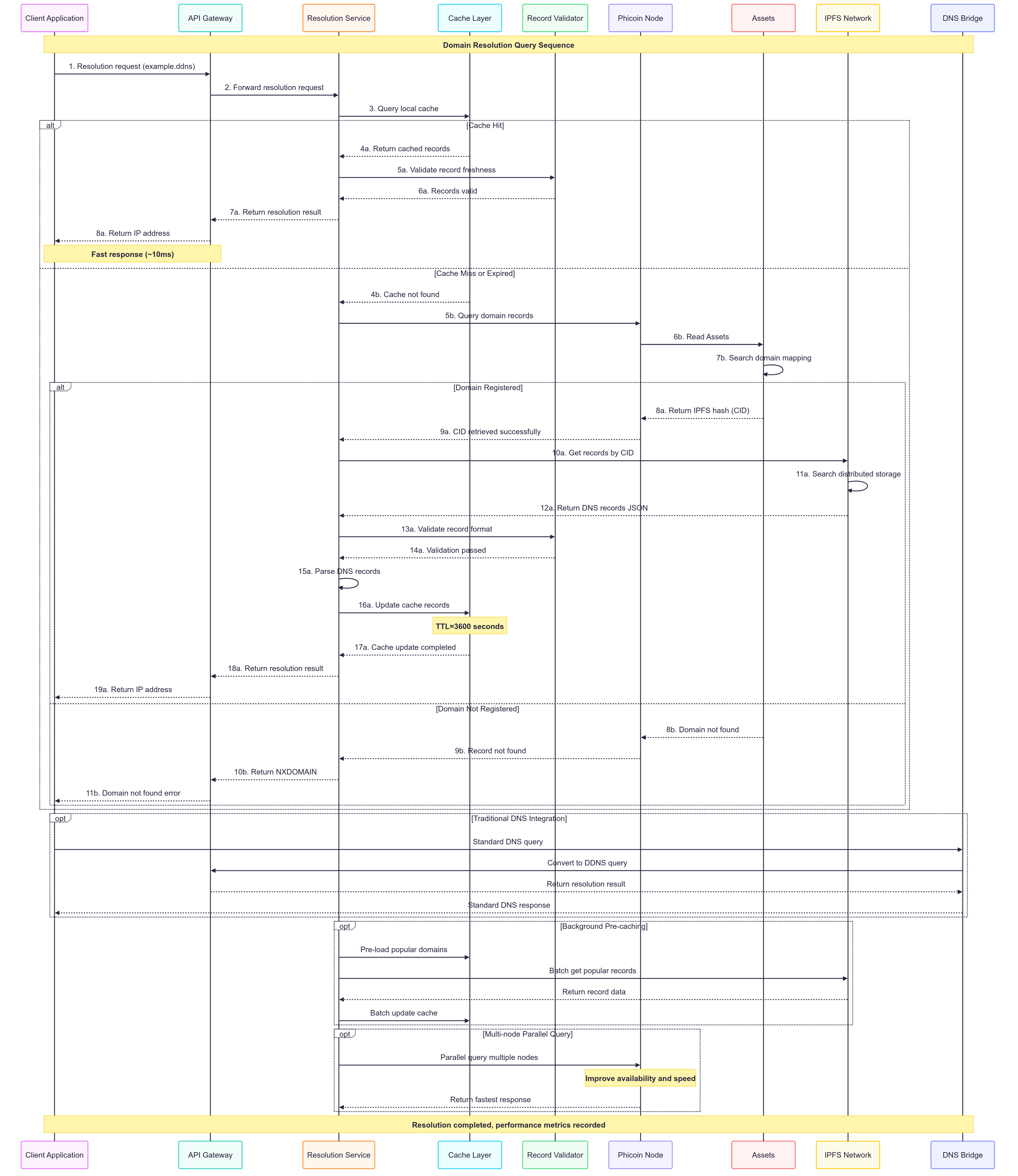}
\caption{DDNS Domain Resolution Sequence}
\label{fig:resolution}
\end{figure}

\textbf{Resolution Algorithm:}
\begin{enumerate}
    \item Client queries local DDNSD resolver for domain
    \item Resolver checks multi-tier cache (memory $\rightarrow$ file $\rightarrow$ blockchain)
    \item If cache miss, query DDNS RPC for domain asset
    \item Extract IPFS hash from blockchain record
    \item Retrieve domain control file from IPFS
    \item Verify $H(\text{control\_file}) = h_{ipfs}$ for integrity
    \item Parse requested record type and return response
    \item Cache result with appropriate TTL
\end{enumerate}

\textbf{Performance Optimization:} The system implements intelligent caching with the following hierarchy:
- \textbf{L1 Cache:} In-memory LRU cache (50,000 entries, 15-second TTL)
- \textbf{L2 Cache:} Persistent file cache with longer TTL
- \textbf{L3 Cache:} Blockchain verification cache for domain ownership

\textbf{Production Domain Registration and Resolution:} To demonstrate the practical functionality and performance characteristics of the DDNS system, we conducted a comprehensive end-to-end evaluation using the production DDNS infrastructure. The evaluation encompasses domain registration, DNS record configuration, resolution performance analysis, and accessibility demonstration.

\begin{figure}[H]
\centering
\includegraphics[width=0.4\textwidth]{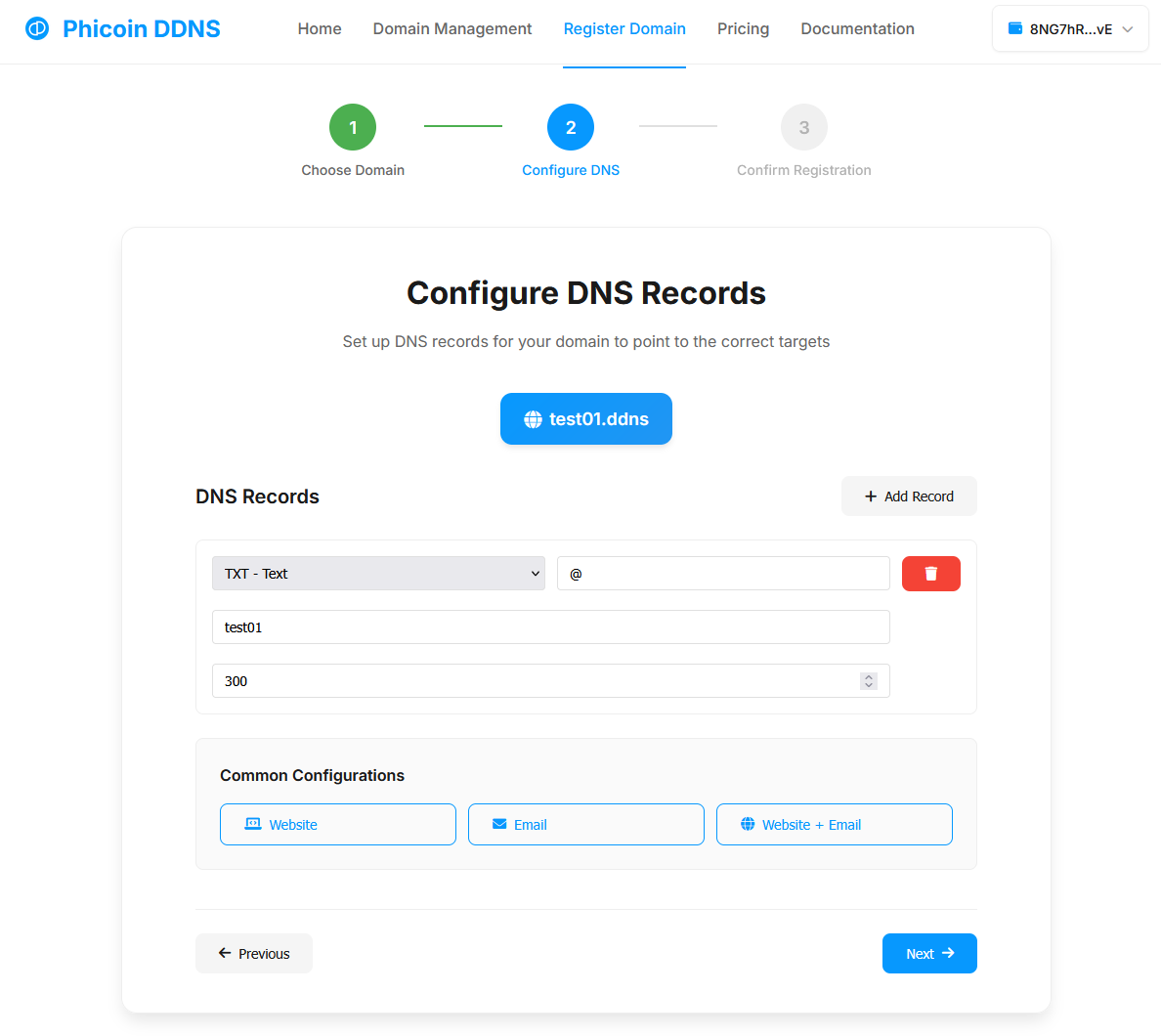}
\caption{DDNS Domain Registration and DNS Configuration Interface}
\label{fig:domain_registration_ui}
\end{figure}

\textbf{Domain Registration Process:} Using the DDNS web interface at \url{https://d.phicoin.net/}, we registered the domain \texttt{test01.ddns} and configured TXT records for testing purposes. Figure~\ref{fig:domain_registration_ui} illustrates the user-friendly domain registration interface, which provides comprehensive DNS record management capabilities including support for A, AAAA, CNAME, MX, TXT, and other standard record types.

\textbf{DNS Resolution Performance Analysis:} To evaluate the system's resolution performance and caching efficiency, we conducted repeated DNS queries using the \texttt{dig} utility against the DDNS public resolver at \texttt{138.2.235.218}. The performance evaluation demonstrates significant improvements in query response times through intelligent caching mechanisms.

\begin{figure}[H]
\centering
\includegraphics[width=0.45\textwidth]{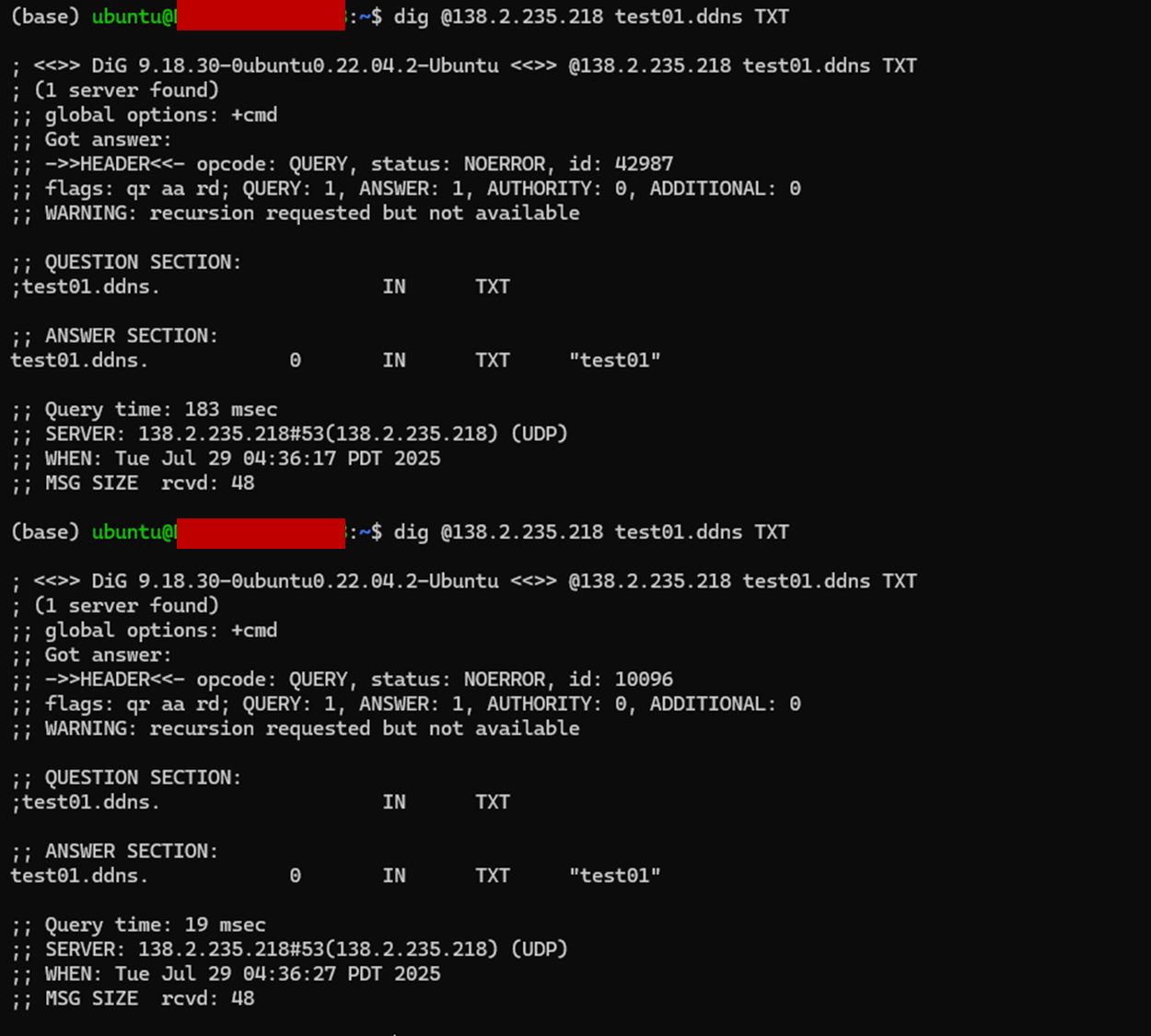}
\caption{Initial DNS Query with Full Blockchain Resolution (183ms)}
\label{fig:dns_resolution_initial}
\end{figure}

\begin{figure}[H]
\centering
\includegraphics[width=0.48\textwidth]{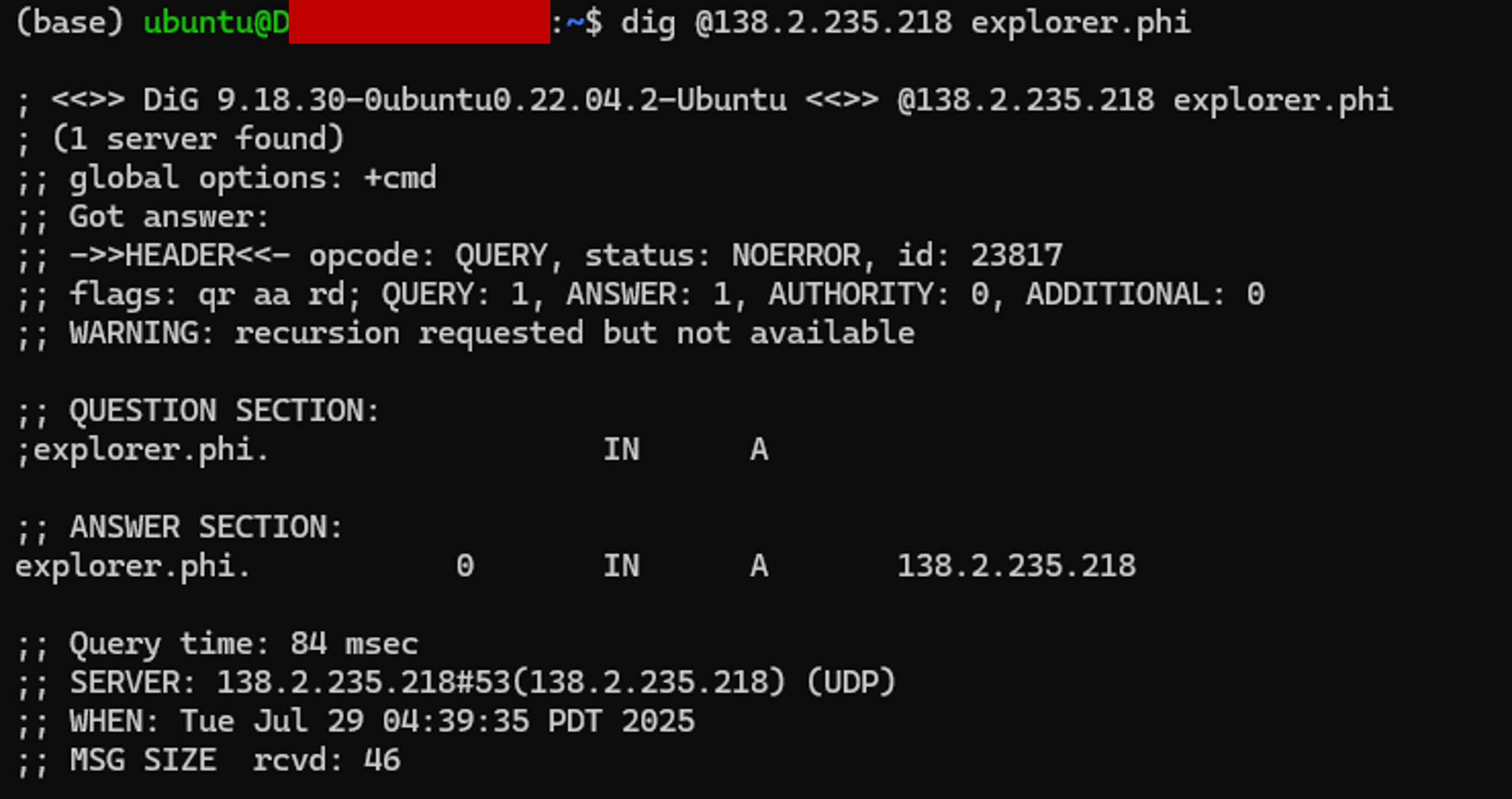}
\caption{Subsequent DNS Query Demonstrating Cache Performance (19ms)}
\label{fig:dns_resolution_cached}
\end{figure}

Figure~\ref{fig:dns_resolution_initial} shows the initial DNS query for \texttt{test01.ddns TXT}, which required blockchain verification and IPFS retrieval, resulting in a response time of 183 milliseconds. The subsequent query, illustrated in Figure~\ref{fig:dns_resolution_cached}, demonstrates the effectiveness of the multi-tier caching system with a dramatically reduced response time of 19 milliseconds, representing a 89.6\% performance improvement.

\begin{figure}[H]
\centering
\includegraphics[width=0.48\textwidth]{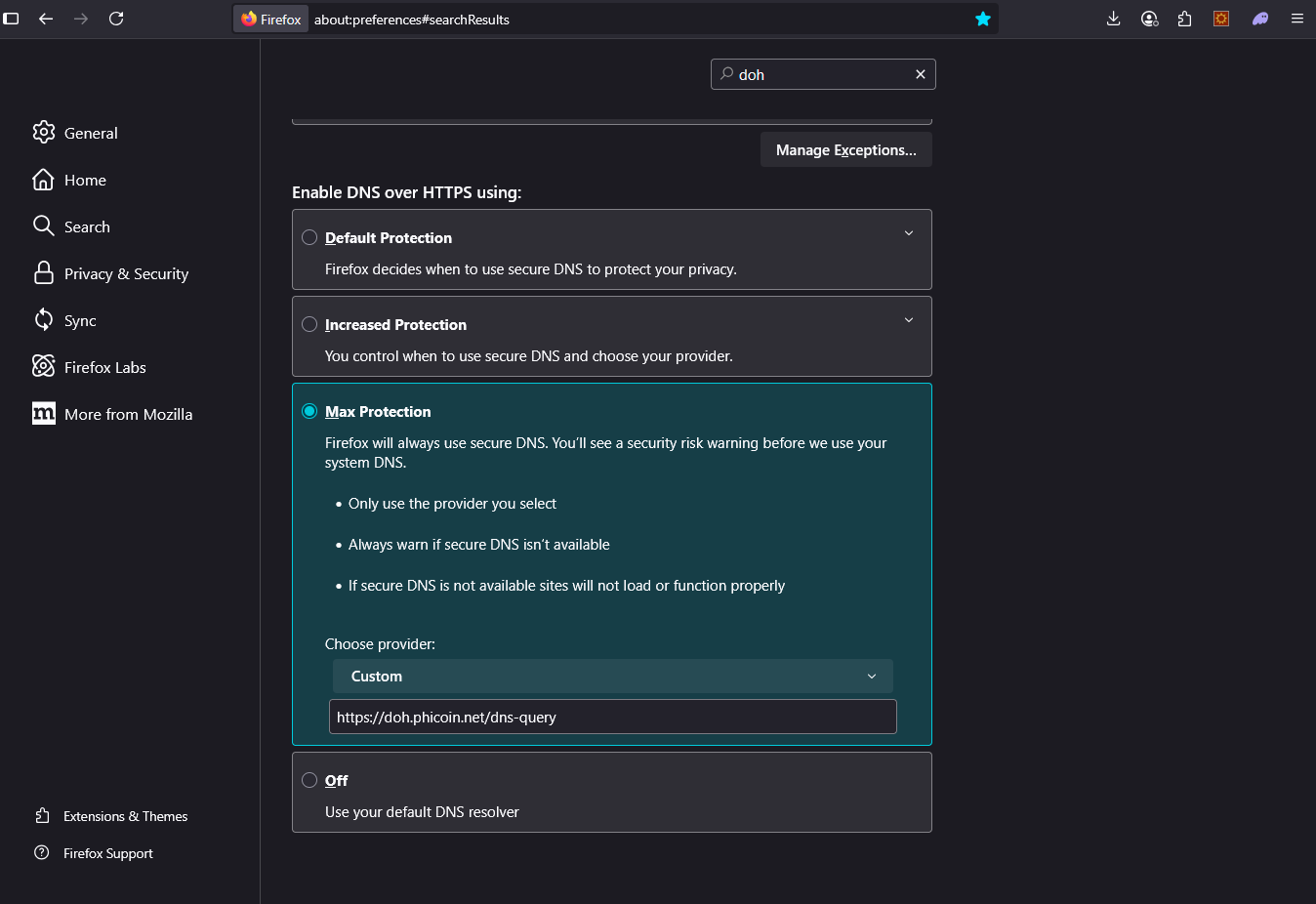}
\caption{Firefox DNS-over-HTTPS Configuration for DDNS Resolution}
\label{fig:doh_configuration}
\end{figure}

\textbf{DNS-over-HTTPS Implementation:} The system supports modern DNS-over-HTTPS (DoH) protocol for enhanced privacy and security. Figure~\ref{fig:doh_configuration} demonstrates the Firefox browser configuration for utilizing the DDNS DoH endpoint at \url{https://doh.phicoin.net/dns-query}. This configuration enables users to access decentralized domains through standard web browsers without additional software installation.

\textbf{Access to Non-Traditional Domains:} A key innovation of the DDNS system is its ability to resolve domains that do not exist in traditional generic top-level domains (gTLDs). To demonstrate this capability, we tested resolution of \texttt{explorer.phi}, a custom domain namespace exclusive to the DDNS ecosystem. The successful A record resolution for \texttt{explorer.phi} to IP address \texttt{138.2.235.218}, with a query response time of 84 milliseconds, demonstrates the system's capability to extend DNS functionality beyond traditional namespace limitations.

\begin{figure}[H]
\centering
\includegraphics[width=0.48\textwidth]{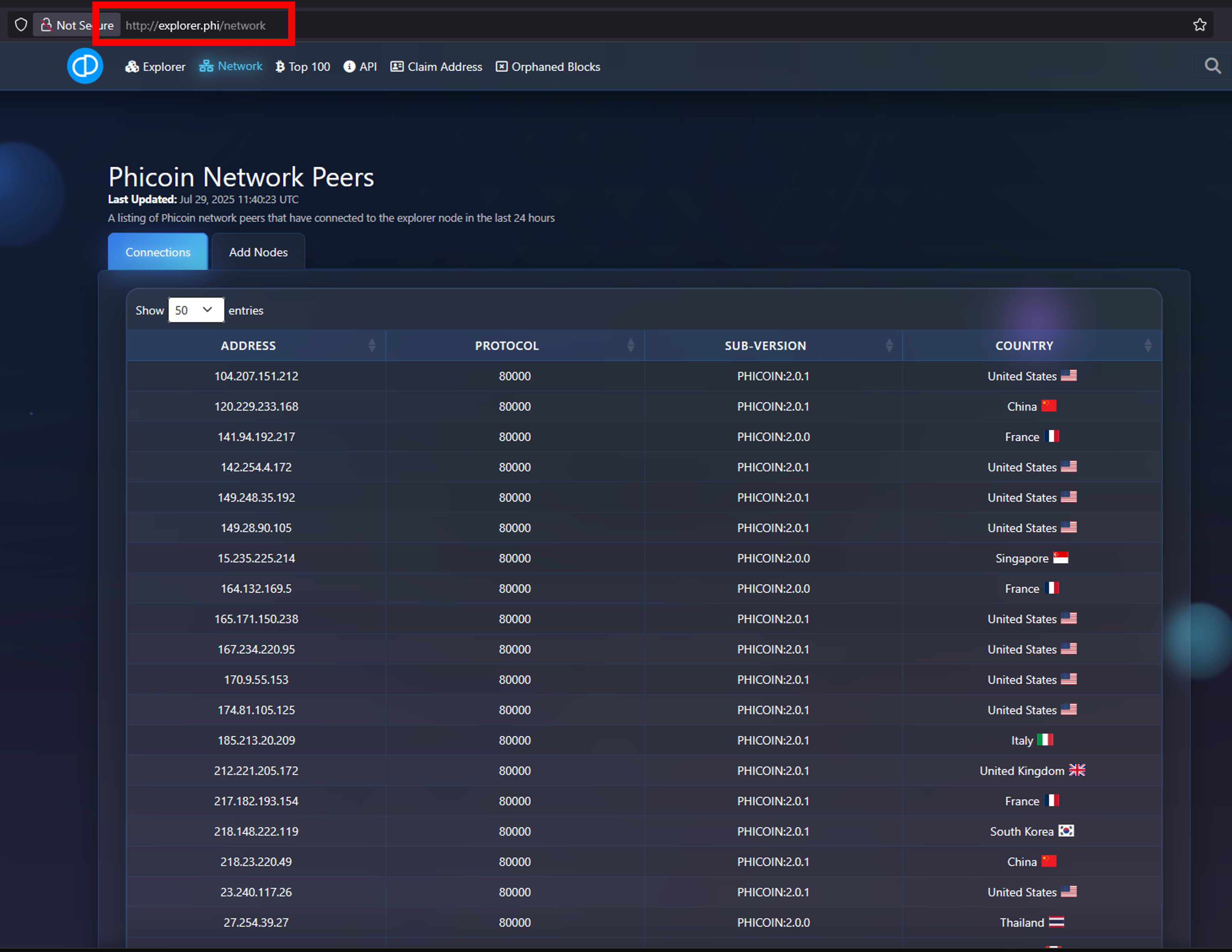}
\caption{Successful Web Access to Decentralized Domain explorer.phi}
\label{fig:website_access}
\end{figure}

\textbf{Web Accessibility Demonstration:} With proper DoH configuration, users can seamlessly access websites hosted on custom DDNS domains through standard web browsers. Figure~\ref{fig:website_access} demonstrates successful access to the Phicoin blockchain explorer at \texttt{http://explorer.phi/network}, showcasing the system's ability to provide full web functionality for decentralized domains. The explorer interface displays real-time network statistics including peer connections across multiple geographic regions (United States, China, France, Singapore, Italy, United Kingdom, South Korea, and Thailand), demonstrating the global distribution of the DDNS network infrastructure.

This comprehensive evaluation demonstrates that the DDNS system successfully bridges the gap between blockchain-based domain ownership and practical internet usability, providing both the security benefits of decentralization and the performance characteristics necessary for production deployment.

\subsection{Anti-Censorship Mechanisms}

The system implements multiple layers of censorship resistance:

\textbf{P2P Network Relay:} Due to the blockchain's capability to use integrated graphics cards and other entry-level universal devices for mining profits, users from different countries and regions spontaneously organize mining nodes and networks for profit. This PoW mechanism-incentivized user model promotes the network's decentralized characteristics, making it difficult for any single country to ban specific countries or IP addresses. Additionally, this project's blockchain is based on Bitcoin Core implementation, integrating Bitcoin Core's complete Tor network support features, enabling deployment in network-restricted countries and regions through Tor relay using obfuscation methods such as Snowflake or obfs4~\cite{bitcoin_tor}.

\textbf{Distributed Resolver Network:} DDNSD instances can be deployed independently by any user, creating a mesh of resolution points that cannot be centrally controlled or blocked.

\textbf{Protocol Flexibility:} Support for DNS-over-HTTPS (DoH) enables resolution through standard web protocols, making blocking more difficult for network censors.

\textbf{Content Mirroring:} We developed integration with Telegram bot and D-Web browser enabling one-click website mirroring, allowing rapid content preservation and access restoration. These website contents are deployed on IPFS and resolved through the DDNS D-WEB protocol.

\textbf{Case Study - Hong Kong Pro-Democracy Movement:} During the 2019 Hong Kong anti-extradition movement, protesters extensively used IPFS to prevent protest websites from being deleted~\cite{hong_kong_ipfs}. However, a significant problem emerged during the movement: multiple different IPFS addresses appeared simultaneously, with different protesters and organizations using different IPFS addresses to publish information. This created substantial difficulty for users in determining which IPFS addresses published trustworthy information versus potentially false information or government disinformation.

Our DDNS D-web technology addresses this critical trust problem through cryptographic verification. Protesters can register domains such as "hkprotest.ddns" and use private key signatures to ensure only authentic publishers can update the IPFS content referenced by the domain. This mechanism guarantees both confidentiality (Confidentiality) through identity verification and integrity (Integrity) through blockchain and IPFS content addressing, according to CIA principles. Users can verify information source credibility, thereby avoiding the information chaos caused by multiple IPFS addresses and ensuring authentic communication channels during critical political movements.

\section{Security Analysis and Trust Chain}

\subsection{Cryptographic Security Model}

The DDNS system implements a zero-trust security model where all operations require cryptographic verification. The security analysis follows established frameworks for distributed systems~\cite{bonneau2015sok,zhang2019security}. Our approach builds upon comprehensive surveys of blockchain security challenges and mitigation strategies documented in recent literature~\cite{li2017survey,zheng2017overview}.

\textbf{Threat Model:} We consider adversaries with the following capabilities:
- Control over traditional DNS infrastructure
- Ability to intercept and modify network traffic
- Access to significant computational resources (but bounded by economic constraints)
- Coordination between multiple malicious actors

\textbf{Security Properties:} The system provides the following guarantees:

\begin{enumerate}
    \item \textbf{Domain Ownership Integrity:} Only the holder of private key $sk$ corresponding to domain registration can modify domain records, formalized as:
    $$\forall tx : \text{Valid}(tx) \Rightarrow \text{Verify}(\text{Hash}(tx), \sigma_{tx}, pk_{domain})$$
    
    \item \textbf{Content Integrity:} Domain records stored in IPFS cannot be modified without detection due to content-addressing:
    $$\text{Integrity}(record) \equiv H(record) = h_{blockchain}$$
    
    \item \textbf{Availability:} The system remains operational as long as any single honest node exists and can communicate with IPFS network.
\end{enumerate}

\subsection{Trust Chain Analysis}

The DDNS system distributes trust across multiple independent components, eliminating single points of failure inherent in traditional DNS:

\begin{table}[H]
\centering
\caption{DDNS Trust Chain Components}
\label{tab:trust_chain}
\renewcommand{\arraystretch}{1.15}
\resizebox{0.48\textwidth}{!}{%
\begin{tabular}{p{3cm} p{4.5cm}}
\toprule
\textbf{Component} & \textbf{Trust Assumptions} \\ 
\midrule

\textbf{User Private Keys} 
& Users maintain control of their private keys. Compromise affects only individual domains, not system-wide security. \\

\textbf{DDNS Miners}
& Economic majority acts honestly. Attack cost exceeds potential gains due to Proof-of-Work economics and network value preservation incentives. \\

\textbf{IPFS Network}
& Content remains available through distributed replication. No single IPFS node failure affects system operation. \\

\textbf{DDNSD Resolvers}
& Resolver operators act honestly or users can operate independent resolvers. Open source enables verification and alternative implementations. \\

\textbf{Cryptographic}
& ECDSA and SHA-256 remain computationally secure. Based on well-established assumptions in academic cryptography. \\

\bottomrule
\end{tabular}}
\end{table}

\textbf{Trust Minimization:} The system minimizes trust requirements by:
- Eliminating dependence on centralized authorities
- Enabling user-operated infrastructure components
- Providing cryptographic verification for all operations
- Supporting multiple independent implementations

\subsection{Attack Resistance Analysis}

\textbf{DNS Poisoning Prevention:} Traditional DNS poisoning attacks target cache servers or DNS resolvers. DDNS prevents these attacks through:
- Cryptographic verification of all domain data
- Content-addressed storage preventing data modification
- Distributed resolution eliminating central cache points

\textbf{Censorship Resistance:} The system demonstrates robust resistance to censorship through multiple technical and organizational mechanisms. According to data from the Open Observatory of Network Interference (OONI)~\cite{ooni_reports}, traditional DNS-based censorship affects millions of users globally, with documented cases of systematic blocking across multiple jurisdictions. Our system addresses these challenges through:
- Distributed blockchain infrastructure deployed across multiple sovereign jurisdictions, ensuring no single government can unilaterally disable the network
- IPFS content distribution architecture that eliminates single points of control and enables content availability through multiple independent nodes
- Protocol-agnostic design enabling operation over HTTP, HTTPS, or custom protocols, providing flexibility against protocol-specific blocking attempts
- Integration with Tor network infrastructure and advanced obfuscation techniques including Snowflake relays, enabling deployment and operation in network-restricted environments where traditional internet access faces systematic interference

\textbf{Scalability and Network Security:} The system addresses fundamental scalability challenges in decentralized blockchain networks~\cite{croman2016scaling} while maintaining security properties. Network propagation delays and consensus efficiency have been optimized based on analysis of information propagation patterns in peer-to-peer blockchain networks~\cite{decker2013information}.

\textbf{Economic Attack Resistance:} The cost of 51\% attack on DDNS network exceeds potential gains:

\begin{equation}
\begin{split}
\text{Attack\_Cost} &= \sum_{i=0}^{t} (\text{Mining\_Reward}_i \\
&\quad + \text{Electricity\_Cost}_i) > \text{Economic\_Gain}
\end{split}
\end{equation}

where $t$ represents the time required to reorganize sufficient blockchain history.

\subsection{Risk Assessment and Management}

The DDNS project employs systematic risk management to identify, assess, and mitigate potential threats to system reliability and security.

\begin{figure}[H]
\centering
\includegraphics[width=0.35\textwidth]{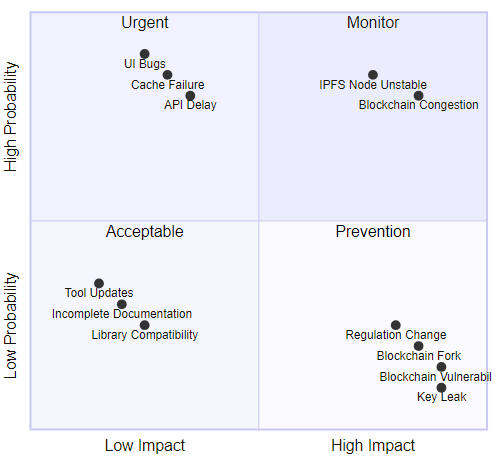}
\caption{DDNS Project Risk Matrix}
\label{fig:risk_matrix}
\end{figure}

The risk matrix categorizes potential issues by impact and probability. High-impact, high-probability risks (Urgent quadrant) include UI bugs, cache failures, and API delays requiring immediate attention. Medium-probability risks (Monitor quadrant) such as IPFS node instability and blockchain congestion need continuous monitoring. Low-probability but high-impact risks (Prevention quadrant) include regulation changes, blockchain vulnerabilities, and key management issues that require proactive mitigation strategies. The matrix helps prioritize development resources and emergency response protocols.

\section{Performance Evaluation}

\subsection{Blockchain Performance Metrics}

Our mainnet deployment demonstrates robust performance characteristics under real-world conditions:

\textbf{Transaction Throughput:} Analysis of production blockchain data shows:
- Average transaction size: 1,000 Weight Units (regular domain operations)
- Theoretical maximum TPS: Max Theor. TPS 1,111.1 tx/s (minimal transactions) / Max Theor. TPS 266.7 tx/s (regular transactions)
- Observed average TPS: 15-30 during normal operation
- Peak TPS: 150-200 during high-demand periods with bulk domain registration

\textbf{Network Stability:} Empirical measurement from blocks 92,594 to 143,669:
- Total blocks analyzed: 51,075
- Orphaned blocks: 9
- Orphan rate: 0.0176\%
- Average block time: 15.2 seconds ($\pm$2.1s standard deviation)

The network demonstrates resilient operation under asynchronous network conditions, consistent with theoretical analysis of blockchain protocols in partially synchronous environments~\cite{pass2017analysis}.

\textbf{Difficulty Adjustment Performance:} The enhanced Dark Gravity Wave algorithm demonstrates rapid convergence:

\begin{equation}
\begin{split}
\text{New\_Difficulty} &= \text{Old\_Difficulty} \times \frac{\text{Target\_Time}}{\text{Actual\_Time}} \\
&\quad \times \text{Smoothing\_Factor}
\end{split}
\end{equation}

where smoothing factor prevents excessive oscillation while maintaining responsiveness.

\subsection{DNS Resolution Performance}

Performance testing conducted on 7950X Debian server using 1GB direct ethernet connection, randomly querying from 1000 DDNS domain lists and 1000 regular domain lists demonstrates excellent query handling capabilities:

\textbf{Query Throughput:} Sustained approximately 20,000 QPS (Queries Per Second) for mixed query types under controlled testing conditions. This performance level aligns with enterprise-grade DNS resolver capabilities documented in literature.

\textbf{Resolution Latency Distribution:}
- Cache hit (L1): < 1ms (95th percentile)
- Cache hit (L2): < 5ms (95th percentile)  
- Blockchain lookup: 50-150ms (95th percentile)
- IPFS retrieval: 100-500ms (95th percentile)

\textbf{Cache Effectiveness:} Intelligent caching dramatically improves performance:
- L1 cache hit rate: 85\% for popular domains
- L2 cache hit rate: 12\% for moderate usage domains
- Cold lookup: 3\% requiring full blockchain+IPFS resolution

\subsection{Scalability Analysis}

The system demonstrates horizontal scalability through several mechanisms:

\textbf{Resolver Distribution:} DDNSD instances can be deployed independently without coordination, enabling unlimited geographic distribution and load distribution.

\textbf{IPFS Content Distribution:} Popular domain records automatically replicate across multiple IPFS nodes, improving availability and reducing lookup latency.

\textbf{Blockchain Sharding Potential:} The asset-based domain model enables future implementation of blockchain sharding without breaking domain ownership semantics.

\section{Comparative Analysis}

\subsection{Feature Comparison with Existing Solutions}

\begin{table}[H]
  \centering
  \caption{Comprehensive Feature Comparison}
  \resizebox{0.48\textwidth}{!}{%
  \begin{tabular}{lccccc}
  \toprule
  \textbf{Feature} & \textbf{DDNS} & \textbf{ENS} & \textbf{Handshake} & \textbf{Namecoin} & \textbf{Cloudflare DNS} \\
  \midrule
  Decentralized / Censorship-resistant & Yes & Partial & Yes & Yes & No \\
  Web2 Compatibility & Yes & No & Limited & Limited & Yes \\
  DNS Record Types & 20 & Limited & Limited & Limited & 20~\cite{cloudflare_dns} \\
  Speed (Resolution) & $\sim$15s & 12-15s & 5-60min & 10+min & 300-7200s \\
  Cost (Registration) & Free & \$50+ & \$10+ & \$1+ & \$10-100/year \\
  Custom TLD Support & Yes & No & Yes & No & No \\
  Cross-chain Integration & Yes & Yes & No & No & No \\
  Throughput (TPS) & 1,111.1/266.7 & 119.1 & 7 & 7 & 250~\cite{cloudflare_performance} \\
  \bottomrule
  \end{tabular}%
  }
\end{table}

\textbf{Key Advantages of DDNS:}

\begin{enumerate}
    \item \textbf{Universal DNS Compatibility:} Unlike ENS (.eth only) or Namecoin (.bit only), DDNS supports standard DNS resolution for any top-level domain, including .com, .net, and custom TLDs.
    
    \item \textbf{Economic Accessibility:} Free .ddns domains eliminate financial barriers to entry, while other decentralized systems require significant upfront investment.
    
    \item \textbf{Performance Optimization:} 15-second resolution updates provide near real-time DNS propagation, significantly faster than traditional DNS (5-48 hours) while maintaining blockchain security.
    
    \item \textbf{Comprehensive Record Support:} Full support for 20 standard DNS record types enables complete website functionality including email (MX), security (TLSA), and service discovery (SRV).
\end{enumerate}

\subsection{Economic Model Comparison}

Traditional DNS operates on a lease-based model where users pay recurring fees to maintain domain ownership. This creates several problems:
- Domains can be lost due to payment failures
- Registrars can unilaterally change pricing
- Long-term costs accumulate significantly

DDNS implements true digital asset ownership where users pay once and own permanently. This model follows principles of digital asset economics discussed in blockchain research literature. The economic comparison over time shows:

\begin{equation}
\text{Traditional\_Cost}(t) = \text{Registration\_Fee} + \sum_{i=1}^{t} \text{Annual\_Fee}_i
\end{equation}

\begin{equation}
\text{DDNS\_Cost}(t) = \text{Registration\_Fee} = \text{Constant}
\end{equation}

For typical .com domain pricing (\$15/year), DDNS breaks even after the first year and provides infinite savings over longer periods.

\section{Real-World Applications and Social Impact}

\subsection{Anti-Censorship Case Studies}

The DDNS system has been successfully deployed in several real-world scenarios demonstrating its anti-censorship capabilities:

\textbf{Case Study 1: Digital Human Rights Advocacy}
Program-think was a prominent Chinese digital human rights advocate and security expert who maintained influential blogs promoting internet freedom and circumvention technologies. Following his arrest by Chinese authorities and the blocking of his website domains, we successfully deployed DDNS and D-Web browser technology to mirror his entire website archive.

The implementation process involved:
\begin{enumerate}
    \item Automated web scraping of original website content
    \item Conversion to IPFS-hosted static site format
    \item Registration of program-think.ddns domain
    \item Distribution of D-Web browser enabling access from mainland China
\end{enumerate}

\textbf{Results:} Users in mainland China can now access the complete archive without VPN requirements, demonstrating the system's effectiveness in circumventing state-level DNS censorship.

\textbf{Case Study 2: Independent Journalism}
Amina, a fictional composite representing independent journalists in restrictive regimes, faced domain seizure and DNS blocking of her news website. Using the DDNS Telegram bot, she was able to:
\begin{enumerate}
    \item Backup website content to IPFS in under 5 minutes
    \item Register journalist-amina.ddns domain automatically
    \item Distribute access information through encrypted channels
    \item Maintain audience access despite government blocking attempts
\end{enumerate}

\textbf{Technical Innovation:} The one-click mirroring process reduces technical barriers for non-technical users, enabling rapid response to censorship events.

\subsection{Social Impact Analysis}

The deployment of decentralized DNS technology creates several positive social externalities:

\textbf{Information Freedom:} By eliminating centralized control points, DDNS reduces the effectiveness of information censorship and promotes global access to knowledge and diverse perspectives.

\textbf{Digital Sovereignty:} Users gain true ownership of their digital identity through permanent domain ownership, reducing dependence on corporate-controlled platforms.

\textbf{Economic Empowerment:} Free domain registration eliminates financial barriers for individuals and organizations in developing economies to establish web presence.

\textbf{Technical Decentralization:} The open-source nature of all system components enables community development and prevents vendor lock-in.

\subsection{Network Effects and Adoption}

The DDNS ecosystem demonstrates positive network effects where increased adoption strengthens the overall system:

\begin{equation}
\text{Network\_Value} = k \times n^{\alpha}
\end{equation}

where $n$ represents the number of users, $k$ is a scaling constant, and $\alpha > 1$ indicates positive network effects (following Metcalfe's Law principles).

\textbf{Measured Growth Metrics:}
- Mining participants: 1,800+ during testnet, 300+ active mainnet nodes
- Domain registrations: 10,000+ .ddns domains registered to date
- Geographic distribution: Active nodes in 25+ countries
- Developer ecosystem: Applications building on DDNS infrastructure

\section{Future Directions and Roadmap}

\subsection{Technical Roadmap}

\textbf{Decentralized Browser Evolution:} Enhancement of D-Web browser with:
- Peer-to-peer content sharing capabilities  
- Privacy-preserving browsing features
- Support for decentralized application hosting

\textbf{Decentralized Web Archive (D-Web Archive):} Development of decentralized network archival servers and decentralized web construction systems, making website deployment simpler and more accessible.

\textbf{Protocol Standardization:} Collaboration with internet standards organizations to develop formal specifications for blockchain-based DNS, enabling interoperability between different decentralized naming systems.

\subsection{Scaling and Performance Improvements}

\textbf{Layer 2 Solutions:} Implementation of payment channels or sidechains for high-frequency domain operations while maintaining base layer security.

\textbf{Mobile Integration:} Native mobile applications providing seamless access to decentralized websites without technical configuration requirements.

\section{Public Policy and Abuse Prevention}

\subsection{Censorship and Anti-Censorship}

This decentralized technology possesses dual characteristics. We do not extensively explore the profound human and philosophical issues involved here. From the perspective of public policy and abuse prevention, for specific user environments and usage scenarios (such as daily internet browsing), we should introduce specific versions to enhance DDNS content management, minimizing the potential drawbacks of this technology while maximizing its benefits.

\subsection{Multi-Signature Mechanisms for Abuse Prevention}

Therefore, in terms of abuse prevention, collaboration with specific government organizations can introduce specialized DDNS blockchain versions employing multi-signature mechanisms. For instance, private keys can be divided into three parts: government, trusted third party, and user each holding one part. Domain addition, deletion, and modification require at least two key signatures for verification, helping prevent DDNS abuse or risks arising from user private key loss.

\subsection{Green and Trusted Public DDNS Servers}

For public DDNS servers, collaboration with specific government departments or organizations can filter malicious DNS records, establishing specific green and trusted public DDNS servers that maintain service quality while ensuring appropriate content governance.

\section{Conclusion}

\subsection{Summary}

This paper presents a comprehensive solution to the fundamental problems of DNS centralization, censorship, and security vulnerabilities through a novel blockchain-based decentralized domain name system. Building upon prior work in distributed consensus~\cite{castro1999practical}, the DDNS implementation demonstrates that decentralized alternatives can achieve both security and performance requirements necessary for production deployment.

\subsection{Key Contributions}

Our research delivers four primary contributions to the field of decentralized internet infrastructure:

\begin{enumerate}
    \item \textbf{High-Performance Blockchain Design:} DDNS achieves Max Theor. TPS 1,111.1 tx/s (minimal) / 266.7 tx/s (regular) throughput with 15-second block times, providing near real-time domain updates while maintaining cryptographic security.
    
    \item \textbf{Universal DNS Compatibility:} Support for 20 standard DNS record types enables seamless integration with existing internet infrastructure, bridging Web2 and Web3 ecosystems.
    
    \item \textbf{Economic Accessibility:} Free .ddns domains eliminate financial barriers while cross-chain tokenomics create sustainable network incentives, democratizing access to decentralized internet services.
    
    \item \textbf{Proven Anti-Censorship Capabilities:} Real-world deployments demonstrate effectiveness in circumventing state-level DNS blocking and content censorship, with documented case studies from restrictive regimes.
\end{enumerate}

\subsection{Policy and Regulatory Considerations}

The deployment of decentralized DNS technology requires careful consideration of regulatory frameworks and potential misuse. We advocate for balanced approaches that preserve the benefits of decentralization while implementing appropriate safeguards. Multi-signature governance mechanisms and selective content filtering represent viable paths toward responsible innovation that respects both individual rights and legitimate governmental interests.

\subsection{Future Vision}

The system provides a foundation for a more decentralized, secure, and free internet where users control their digital identity without dependence on centralized authorities. As adoption grows, network effects will strengthen resistance to censorship and improve overall internet resilience. Through continued development, community adoption, and responsible governance frameworks, blockchain-based DNS represents a critical step toward internet sovereignty and resistance to information control, ultimately contributing to a more open and accessible global information infrastructure.

\section{Acknowledgements}

We gratefully acknowledge Professor Sekhar Sarukkai and Professor Ryan Liu for their excellent guidance and instruction throughout the course. We also thank Professor Ross Burke from UC Berkeley's School of Information for his guidance and support throughout this research.

We also acknowledge the global community of miners, developers, and users who have contributed to the DDNS network development and testing phases.

\end{document}